\documentclass[nohyper,12pt,letterpaper]{JHEP}  
\usepackage{epsfig}
\newfont{\frak}{eufm10 scaled 1200}

\newfont{\Bbb}{msbm10 scaled 1200}     %instead of eusb10
\newcommand{\mathbb}[1]{\mbox{\Bbb #1}}
\newcommand{\be}{\begin{eqnarray}}
\newcommand{\ee}{\end{eqnarray}}

\newcommand{\f}{\phi}

\newcommand{\e}{\varepsilon}

\newcommand{\p}{\partial}

\newcommand{\cN}{{\cal N}}

\renewcommand{\p}{\partial}

\DeclareSymbolFont{AMSa}{U}{msa}{m}{n}
\DeclareSymbolFont{AMSb}{U}{msb}{m}{n}
\let\Box\relax
\DeclareMathSymbol{\Box}{\mathord}{AMSa}{"03}

\title{Subleading Corrections and Central Charges in the
AdS/CFT Correspondence}

\author{D. Anselmi and A. Kehagias \\
  Theory Division, CERN\\
  1211 Geneva 23, Switzerland\\
E-mails: \email{Damiano.Anselmi@cern.ch},~~
\email{Alexandros.Kehagias@cern.ch}}

\abstract{We explore subleading contributions to the two basic central 
charges $c$ and $a$ of four-dimensional conformal field theories in the AdS/CFT
scheme. In particular we probe subleading corrections to the 
difference $c-a$ from the string-theory side. In the $\cN=4$ CFT, $c-a$ vanish
identically consistently with the string-theory expectations. However,
for $\cN=1$ and $\cN=2$ CFTs, the  $U_R(1)$ anomaly, which is 
proportional to $c-a$, is subleading in the large $N$ limit  
for theories in the AdS/CFT context
and one expects string one-loop 
$R^2$ and $B\wedge R\wedge R$ terms in the low energy effective action. 
We identify these terms as coming from the $R^4$ terms. 
 Similar considerations  apply
to the  $U_R(1)^3$ anomaly which is, however, subleading only 
for $\cN=2$ theories. As a result, a string one-loop term $B\wedge F\wedge F$
 should exist in the low energy effective action of the $\cN=4$ 
five-dimensional supergravity. The  $U_R(1)^3$ term 
is leading for the $\cN=1$ CFT 
and it is indeed present in the $\cN=2$ five-dimensional supergravity.
}


\received{???????? ?st, 1998}
\accepted{???????? ?th, 1998}

\preprint{\hepth{9812092}\\CERN-TH/98-394}
\begin{document}

%%%%%%%%%%%%%%%%%%%%%%%%%%%%%%%%%%%%%%%%%%%%%%%%%%%%%%%%%%%%%%%%%%%%%%%%%%%%
%          Table of contents automatic !!!                                 %
%%%%%%%%%%%%%%%%%%%%%%%%%%%%%%%%%%%%%%%%%%%%%%%%%%%%%%%%%%%%%%%%%%%%%%%%%%%%

\section{Introduction and Conclusions}
% ==========================================================================

It has recently been argued in \cite{malda}
that the large $N$ limit of certain conformal field theories (CFT)
can be described in terms of Anti de-Sitter (AdS)  
supergravity. The CFT
lives on the AdS boundary and a precise recipe for expressing
correlation functions of the boundary theory in terms of the bulk theory 
has been given \cite{kleb},\cite{wit1}.    
In particular,  the four-dimensional ${\cal N}=4$ supersymmetric 
$SU(N)$ Yang-Mills theory is described by the type IIB string theory  on 
$AdS_5\times S^5$ where the radius of both 
the $AdS_5$ and  $S^5$ are proportional 
to $N$. A field theory formulation of the proposed 
AdS/CFT correspondence has been 
given in \cite{FF},\cite{FZ}.
 It has also been argued that 
in a suitable limit, the generating functional for the boundary 
correlators is reproduced \cite{kleb},\cite{wit1} 
by the maximal $\cN=8$ $d=5$ gauged supergravity
on its anti-de Sitter vacuum \cite{GRW}. The symmetry of the latter  
 is $SO(4,2)\times SU(4)$ which is just the even subgroup of the 
$SU(2,2|4)$ superalgebra. The latter is realized by $\cN=4$ superconformal 
YM theory on the four-dimensional boundary of the anti-de Sitter space.

   In addition to the ${\cal N}=4$ 
supersymmetry algebra $SU(2,2|4)$,
there also exist the superalgebras $SU(2,2|2)$ and 
$SU(2,2|1)$. Their  
even subgroups are  $SO(4,2)\times U(2)$ and $SO(4,2)\times U(1)$, 
respectively, and they are realized by  conformal field theories with less 
supersymmetries, namely, ${\cal N}=2$ and ${\cal N}=1$ superconformal 
Yang-Mills theories. In this case, the boundary correlators are reproduced
by the $\cN=4$ and $\cN=2$ $d=5$ gauged supergravity \cite{Ro},\cite{GT}. 
 However, this way one may explore only   the leading $N^2$ terms since  
classical supergravity arises from tree-level string theory and so there
exist 
a $1/g_s^2$ factor in front of its effective action. Recalling that
 $g_s\sim1/N$, we immediately conclude that classical supergravity 
is of order $N^2$. Thus, in order to probe the subleading structure, one 
has to go beyond tree-level string theory and  take into account 
string-loop effects. Here we will explore subleading contributions to 
the basic central charges of a four dimensional
conformal theory, commonly called $c$ and $a$ \cite{noi}. 
Quantum field theoretical knowledge is used as 
a guideline to identify the desired terms in the string description.
The leading contributions to $c$ and $a$ have been calculated, in the 
holographic context, in refs. \cite{kleb},\cite{HS}.

One may ask if the field-theory/string-theory correspondence
can be extended so that for any given $\cN=0,1,2$ superconformal 
model in four dimensions there exist a supergravity theory on $AdS_5$. It 
seems, however,  from the known examples discussed so far \cite{Gu}
that  the correspondence works only when $c=a$
at the leading order.
We may conjecture then that {\it all CFT with c=a in the leading order 
have a supergravity dual}. This is also indicated from the present work 
on subleading corrections. 
Conformal field theories with $c=a$ are a special subclass of 
the more general family with $c$ and $a$ unconstrained \cite{n=2}, \cite{anom}.
This and other features visible from the quantum
field theoretical viewpoint fit nicely with the supergravity description
and our purpose here is to show that they are consistent 
also with the subleading
corrections that have a string origin. 

The first quantity to probe is the difference $c-a$.
Given that $c=a$ in the supergravity limit, the presence
of subleading effects can be detected as a non-vanishing,
subleading, value of the difference $c-a$. 
This contribution appears in
theories with $\cN=1$ and $\cN=2$ supersymmetry and does
not appear in theories with $\cN=4$. In the former cases 
a corresponding term, that we shall discuss in detail, is present 
in the string-theoretical description.

$c-a$ is a multiplicative factor of a four-derivative term 
in field theory. Thus, we expect that these terms  correspond 
to four-derivative 
interactions in five-dimensional supergravity. In addition, the 
latter should be of order ${\cal O}(1)$ compared to the leading
$N^2$ terms. Thus, they should emerge from one-loop in string theory. 
Such string one-loop four-derivative interactions 
in five dimensions 
are induced by $R^4$ terms in ten \cite{GZ},\cite{GW} 
or eleven dimensions \cite{FT},\cite{GV}. Since their structure
depends on the number of supersymmetries as well as on 
the particular compactifications, we will examine these terms separately 
according to the number of supersymmetries.

\section{${\cal R}^4$ terms in string and M-theory}

The massless spectrum of the ten dimensional 
 type II string theory contains in its 
$NS/NS$ sector the graviton $g_{MN}$, the antisymmetric two-form $B_{MN}$ and 
the dilaton $\f$. The $R/R$ sector of the IIA theory contains a one-form $A_M$
and a  three-form $A_{MNP}$ while the $R/R$ sector the 
 type IIB theory  consists of a second scalar $\chi$, a two-form 
$B_{MN}^{(2)}$ and a four-form $A_{MNPQ}^+$ with self-dual field strength. 
 In particular the massless spectrum of the type IIA theory  
can be obtained from the 
dimensional  reduction of the  eleven-dimensional supergravity on a circle. 
In that case, the dilaton is related to the radius of the circle, 
the one-form is the KK potential, and the two-and three-forms result from 
the three form of eleven-dimensional supergravity. 

In the large wavelength limit,  the type IIA and IIB  theories are 
described by the 
 non-chiral and chiral $\cN=2$ supergravity, respectively,
 and the bosonic part of their
low energy effective action of the NS/NS sector is  
\be
S_t=\frac{1}{2\kappa_{10}^2}\int d^{10}x \sqrt{-g}e^{-2\f}\left( R+4\p\f^2
-\frac{1}{12}H_{KMN}H^{KMN} \right)\, , \label{st}
\ee
where $\kappa_{10}^2=2^6\cdot \pi^7\cdot \alpha'^4$. 
The first terms in the effective action that receive
quantum corrections are 
the eight-derivative ones. Such terms  are 
the familiar $t_8t_8R^4, \, \e_{10}\e_{10}R^4 $ and $t_8\e_{10}
R^4$ where 
\be
 t_8t_8R^4&=& t_8^{M_1M_2...M_7M_8}t_8^{N_1N_2...N_7N_8}R_{M_1M_2N_1N_2} ... 
 R_{M_7M_8N_7N_8}\, , \nonumber \\
\e_{10}\e_{10}R^4&=&{\e}^{M_1M_2...M_7M_8AB}
{{\e}^{N_1N_2...N_7N_8}}_{AB}R_{M_1M_2N_1N_2} ... 
 R_{M_7M_8N_7N_8}\, , \nonumber \\
t_8\e_{10}R^4&=&{\e}^{ABM_1M_2...M_7M_8}
t_8^{N_1N_2...N_7N_8}R_{M_1M_2N_1N_2} ... 
 R_{M_7M_8N_7N_8}\, . 
\ee
The eight-tensor  $t_8$ appears in string amplitude calculations \cite{S}
and $\e_{10}$ is the ten-dimensional totally antisymmetric symbol. 
In particular,  using the explicit form of $t_8$ we find \cite{TT},\cite{KP}
\be
  t_8t_8R^4&=&6\, t_8\left(4R^4-(^2TrR^2)\right)\nonumber \\
&=&12 \left(R_{MNPQ}R^{MNPQ}\right)
-192R_{MNPQ}{R^{MNPK}}R_{ABCQ}R^{ABCQ}\ldots \, , \nonumber \\
 \e_{10}\e_{10}R^4&=&-96  \left(R_{MNPQ}R^{MNPQ}\right)\ldots \,  , 
\nonumber 
%\\t_8\e_{10}R^4&=&24 TrR^4-6\left(TrR^2\right)^2.
\ee
The next-to-leading order corrections to the tree effective action  
 can be computed  either by 
string amplitude calculations or in sigma-model perturbation theory. Both 
ways lead to the result that the eight-derivative term in the effective 
action is of the form 
$t_8t_8R^4$. However, 
in the case of ten-dimensional $\cN=1$ 
supergravity, it has been shown that supersymmetry relates the $t_8t_8R^4$
term to $\e_{10}\e_{10}R^4 $ \cite{BR}. In fact what appears in the 
effective action are 
the two super-invariants with bosonic parts
\be
J_0=t_8t_8R^4+\frac{1}{8} \e_{10}\e_{10}R^4\, , ~~~
J_1=t_8X_8-\frac{1}{4} \e_{10}BX_8 \, ,
\ee
where 
\be
X_8=\frac{1}{(2\pi)^4}\left(-\frac{1}{768}(TrR^2)^2+
\frac{1}{196}TrR^4\right)\, ,
\ee
is the eight-form anomaly polynomial. We will assume that that 
this is also the case for the $\cN=2$ supersymmetry 
and then the eight-derivative tree-level effective action turns out to
 be
\be
S^{(0)}_{R^4}=
\frac{1}{3\cdot2^{12}\cdot\kappa_{10}^2}\zeta(3)\int d^{10}x\sqrt{-g}
e^{-2\f}\left(t_8t_8+\frac{1}{8} \e_{10}\e_{10}\right)R^4\, . \label{tree} 
\ee
We see that $J_1$ contains a  CP-odd coupling and thus it is expected to be
protected from perturbative corrections and all possible corrections are
non-perturbative ones. On the other hand, $J_0$ is believed to have
only one-loop corrections. In particular, $J_0$ also appears in type IIB 
theory. There, its perturbative and non-perturbative corrections can be 
extracted from symmetry considerations \cite{GG},  namely, the $SL(2,{\bf Z})$ 
invariance. The latter specifies the form of the corrections not 
only to the $R^4$ term but of all eight-derivative terms \cite{KP1}. 
The result is that there exist only 
one-loop corrections to these terms and the 
non-perturbative ones are due to  the type IIB D-instantons.    
In fact,  $SL(2,{\bf Z})$ symmetry is even stronger. One may prove that 
higher-derivative gravitational interactions \cite{RT}
are the form $R^{6L+4}$ 
($L=0,1,...$) and they appear at $L$ and $2L+1$ loops \cite{KP2}. 
For $L=0$ this is just the statement that the  tree 
level $R^4$ term  has only one-loop counterpart and
all other corrections are 
non-perturbative. 

The one-loop effective action of the type IIA theory  turns out to be 
\be
S^{(1)}_{R^4}=
\frac{2\pi^2}{3}\frac{1}{3\cdot2^{13}\cdot\kappa_{10}^2}\!\!
\int d^{10}x\sqrt{-g}
\left(\!\!\!\!\!\!\!\!\!\!\!\!\phantom{\frac{E^{A^B}}{\frac{1}{X}}}
\!\!\!\left(t_8t_8-\frac{1}{8} \e_{10}\e_{10}\right)R^4\!\!-
\!\!B\wedge\left(\!\!\!\!\!
\phantom{\frac{1}{8}}(TrR^2)^2\!\!-\!\!4TrR^4\right)\right)\, , 
\label{1-loop}
\ee 
which contains a CP-odd and a CP-even part. Let us note at this point that the 
CP-odd term is absent in the type IIB theory since the transformation 
$-I_{2\times2}\in SL(2,{\bf Z})$ changes the sign of the two-forms 
(i.e., $B\rightarrow-B$) and thus the last two-terms in  eq.(\ref{1-loop}) is 
absent. Moreover, the CP-even term is different also 
in type IIB and is proportional to $\left(t_8t_8+\frac{1}{8} 
\e_{10}\e_{10}\right)R^4$. 

By relating the dilaton with the radius $R_{11}$ of an $S^1$ 
compactification of M-theory as $e^\phi=R_{11}^{3/2}$, one may lift the 
tree and one-loop effective actions eqs.(\ref{tree},\ref{1-loop}) to 
eleven-dimensions.  One then finds that in the decompactification limit 
$R_{11}\to \infty$ only the one-loop effective action in eq.(\ref{1-loop})
survives and the result is 
\be
S^{11}_{R^4}=
\frac{2\pi^2}{3}\frac{1}{3\cdot2^{13}\cdot\kappa_{10}^2}\!\!
\int d^{10}x\sqrt{-g}
\left(\!\!\!\!\!\!\!\!\!\!\!\!\phantom{\frac{E^{A^B}}{\frac{1}{X}}}
\!\!\!\left(t_8t_8-\frac{1}{24} \e_{11}\e_{11}\right)R^4\!\!-
\!\!C\wedge\left(\!\!\!\!\!
\phantom{\frac{1}{8}}(TrR^2)^2\!\!-\!\!4TrR^4\right)\right) , 
\label{M}
\ee 
where $C$ is the three-form of eleven-dimensional supergravity. 
In fact, the last term in 
eq.(\ref{M}) is needed to cancel the anomaly on the 
fivebrane world-volume by a bulk contribution \cite{DM}.

\section{${\cal R}^2$ terms in $d=5$ supergravity}

There exist four supergravity theories in five dimensions, the $\cN=8$,
$\cN=6$, 
$\cN=4$ and $\cN=2$ supergravities \cite{C}. 
Since according to the AdS/CFT scheme these theories will correspond 
to supersymmetric $\cN/2$ YM theories, we will only consider 
the $\cN=8$, 
$\cN=4$ and $\cN=2$ $d=5$ theories. They can be obtained by compactifications
of M-theory on $T^6$, $K3\times T^2$ and Calabi-Yau $(CY)$, respectively.
The presence of $R^4$ terms in eleven dimensions yield $R^4$ as well as 
$R^2$ terms in five dimensions after compactification. 
In particular the presence of the latter depends on the number 
of supersymmetries. Namely, $R^2$ terms in five dimensions,
 which are one-loop and thus
subleading with respect to the two-derivative terms, exist,  as we will see, 
only for the  $\cN=4$ and $\cN=2$ case  and not for $\cN=8$. We 
should stress here that the $R^2$ terms we are discussing appear in  
the ungauged theory. The latter have a $USp(\cN)$ group of local symmetries 
  and one may gauge an 
appropriate subgroup of it. In this case, the $R^2$ terms as well as
the $R^4$ terms should also exist in the gauged theory since 
in the limit of vanishing gauge coupling one should recover these terms.
This is also supported from the fact that these terms are needed 
for the consistency of the AdS/CFT correspondence. 
Namely, the one-loop $R^2$ terms which are subleading with respect to 
the two-derivative terms in the supergravity side provide the  necessary
and correct  structure to produce in the CFT side the R-current 
anomalies which are proportional to $c-a$. Note that $c-a$ is exactly zero
for the $\cN=4$ SCFT while it is subleading in the $\cN=1,2$ case. 
Thus, we  expect $R^2$ terms in the supergravity side only for 
the $\cN=2,4$ and not for the $\cN=8$ $d=5$ supergravity. This is indeed 
what we find and supports the fact that the $R^2$  in the ungauged 
theory  also exist in the gauged one.

\subsection{The maximal $\cN=8$ $d=5$ supergravity}

The maximal $\cN=8$ five-dimensional ungauged supergravity theory has been
constructed in \cite{C}. It can be obtained by toroidal 
compactification of M-theory which has the eleven-dimensional supergravity 
as its low-energy limit. It has a graviton, 
eight symplectic Majorana gravitini, 27 vectors, 48 symplectic 
Majorana spinors and 42 scalars.  
It has an $E_{6(6)}$ global and a local $USp(8)$ symmetry. The scalar
fields parametrize the coset space $E_{6(6)}/USp(8)$. An $SO(p,6-p)$ 
$(3\leq p\leq6)$ subgroup of  $E_{6(6)}$ can be gauged 
resulting 
in the maximal gauged supergravity in five dimensions \cite{GRW}. 
In particular, for $SO(6)=SU(4)$ gauging, the supergravity   
 admits an $AdS_5$ vacuum which exhibits the $SU(2,2|4)$ superalgebra
and according to the AdS/CFT correspondence, describes large 
$N$ $SU(N)$ $\cN=4$ YM theory at the boundary of $AdS_5$. Note that 
the maximal  gauged supergravity may have vacua with less 
supersymmetries. However, there is no complete 
classification of the critical points 
of the potential of the $\cN=8$ gauged supergravity \cite{GZP},\cite{DZZ},
\cite{PWW}.    

Since the ungauged  theory can be obtained by toroidal compactification
of M-theory we do not expect four-derivative interactions. The first 
non-zero higher-derivative terms are eight-derivative ones which 
in the AdS/CFT context has been discussed in \cite{BG}.  

\subsection{The  $\cN=4$ $d=5$ supergravity}

The five-dimensional $\cN=4$ supergravity has been constructed in \cite{C}
by truncation of the $\cN=8$ theory and its action has explicitly be 
written in \cite{AT} where its coupling to $n$ vector multiplet has also 
be considered. 
The $\cN=4$ $d=5$ supersymmetry algebra has $USp(4)$ as its automorphism 
group and the graviton multiplet contains six vectors in the $5+1$ rep.
of $USp(4)$. Since the bosonic subgroup of the $\cN=4$ anti-de Sitter 
supergroup $SU(2,2|2)$ is $SU(2,2)\times SU(2)\times U(1)$, an $SU(2)\times
U(1)$ subgroup of  $USp(4)$ can be gauged \cite{Ro}. 

The $\cN=4$ $d=5$ supergravity can be obtained by compactification of 
M-theory on $K3\times T^2$, or equivalently of type IIA and IIB 
on $K3\times S^1$. 
The $R^4$ terms in string or M-theory can potentially give rise to 
$R^2$ terms as well. To find the 
 explicit form of these contributions to the effective action we will
consider first  compactification of the ten-dimensional type IIA theory 
theory on $K3$ and a further compactification on $S^1$. For the $K3$ 
compactification there exist two $R^2$ type of terms.  
The ones coming from the tree level eq.(\ref{tree}) effective action 
$S_{R^4}^{(0)}$  and those coming from the one-loop action $S_{R^4}^{(1)}$
in eq.(\ref{1-loop}). One may easily verify that the former are zero while the 
latter are non-zero and they are given explicitly by
\be
S_{R^2}^{(1)}=
\frac{\pi^2}{3\cdot2^7\cdot\kappa_{10}^2}\int d^{6}x\sqrt{-g}
\left(R_{\bar{m}\bar{n}\bar{p}\bar{q}}
R^{\bar{m}\bar{n}\bar{p}\bar{q}}-\frac{1}{4}
\e^{\bar{m}\bar{n}\bar{p}\bar{q}\bar{r}\bar{s}}
B_{\bar{m}\bar{n}}R_{\bar{a}\bar{b}\bar{p}\bar{q}}R_{\bar{a}\bar{b}\bar{r}
\bar{s}}\right)
\, . \label{11}
\ee
where $\bar{m},\bar{n}\ldots=0,...,5$. 
%&&
%\!\!\!\!\!\!\!\!\!\!\!\
%\left(\!\!\!\!\!\!\!\!\!\!\!\!\phantom{\frac{E^{A^B}}{\frac{1}{X}}}
%\frac{1}{3}\left(t_8t_8-\frac{1}{8} \e_{10}\e_{10}\right)R^4 \right.
%\nonumber \\&&\left.
%\!\!\!\!\!\!\!\!\!\!\!\!\!\!\!\!\!\!\!\!\!\!\!\phantom{\frac{E^{A^B}}
%{\frac{1}{X}}}
%-B\wedge\left(\!\!\!\!\!
%\phantom{\frac{1}{8}}(TrR^2)^2-4TrR^4\right)\right)\, , 
%\label{1-loop}
%\ee 
As a result, the only $R^2$ terms existing in $\cN=2$ six-dimensional theory
are at one-loop level as has also been found by a direct string 
one-loop calculation \cite{KKP}.
This can also be infield from the heterotic/type 
IIA duality according to which heterotic string theory on $T^4$ is dual to 
type IIA theory on $K3$. The tree-level 
effective action of the ten-dimensional heterotic
string has $R^2$ terms which upon reduction on $T^4$ will give similar 
terms in six-dimensions. By using the heterotic-type IIA mapping these terms 
can be written in the dual IIA theory and will become one-loop terms. These 
are the terms we found by direct compactification in the type IIA side on 
$K3$. Let us recall here, that, the compactification of the IIA string on 
$K3$ will give rise to a six-dimensional theory with massless sector consisting
of the graviton and vector multiplets of the non-chiral $(1,1)$ supersymmetry.
In particular, the ten-dimensional graviton will give rise to the 
six-dimensional graviton together with $58$ scalars. 
The two-form $B_{MN}$ will 
give rise to a two-form $B_{\bar{m}\bar{n}}$ and $22$ scalars and 
together with the 
dilaton we get $81$ scalars from the NS/NS sector. On the other hand, 
the R/R sector provides $24$ vectors. Thus, finally, we end up 
with the $(1,1)$ six-dimensional graviton multiplet which contains 
the graviton, 1 antisymmetric two-form,  4 vectors, 1 real scalar, 4 Weyl
spinors and 2 gravitini together with 20 vector multiplets each one 
containing 1 vector, 2 Weyl spinors and 4 scalars. Note that the scalars
parametrize the space $R^+\times SO(4,20)/(SO(4)\times SO(20))$. 
The six-dimensional effective action for the non-chiral 
six-dimensional supergravity with no vectors  has been given 
in \cite{R}. Then, together 
with the one-loop $R^2$ terms of eq.(\ref{11}) we have
\be
S_6=\frac{1}{2\kappa_{6}^2}\int d^{6}x \sqrt{-g}\!\!&&\!
\left(\!\!\!\!\!\!\
\phantom{\frac{{E^A}^A}{{E^A}^A}}\!\!\!\!\!\!\!e^{-2\f}\left( R+4\p\f^2
-\frac{1}{12}H_{\bar{k}\bar{m}\bar{n}}H^{\bar{k}\bar{m}\bar{n}}
\right)\right.\nonumber \\&&
-\frac{1}{2}F^I_{\bar{m}\bar{n}}F_{I}^{\bar{m}\bar{n}}
-\frac{1}{8}\e^{\bar{m}\bar{n}\bar{p}\bar{q}\bar{r}\bar{s}}
B_{\bar{m}\bar{n}}F^I_{\bar{p}\bar{q}}F_{I\bar{r}\bar{s}}\label{sts}
 \\ &&\!\!\!\!\!\!\!
\phantom{\frac{{E^A}^A}{{E^A}^A}}\!\!\!\!\!\!\!
\left.
+\frac{1}{8}\alpha'\left(R_{\bar{m}\bar{n}\bar{p}\bar{q}}
R^{\bar{m}\bar{n}\bar{p}\bar{q}}-\frac{1}{4}
\e^{\bar{m}\bar{n}\bar{p}\bar{q}\bar{r}\bar{s}}
B_{\bar{m}\bar{n}}R_{\bar{a}\bar{b}\bar{p}\bar{q}}R_{\bar{a}\bar{b}\bar{r}
\bar{s}}
\right)\!\!\!\!\!\!\!
\phantom{\frac{{E^A}^A}{{E^A}^A}}\!\!\!\!\!\!\!\right)
\, , \nonumber 
\ee
 where $F^I=dA^I,\, I=1,...,4$ are  the field strengths of 
the four vectors $A^I$ of the graviton multiplet and 
$\kappa_{6}$ is the six-dimensional gravitational coupling constant.

By a further compactification on $S^1$, we get the $\cN=4$ five-dimensional
theory. In this case,  the six-dimensional graviton multiplet 
yields the  five dimensional graviton multiplet and a 
vector multiplet. In particular, the six-dimensional 
graviton will give rise to the graviton, one vector and one scalar while
$B_{\bar{m}\bar{n}}$ 
will give rise to a vector $B_m=B_{m5}$ $(m=0,...,4)$ and 
an antisymmetric two-form which can be dualized 
to a vector as well. The four vectors will result into four vectors and 
four scalars and in addition we will have one more scalar $\phi$. The graviton 
together with $5+1$ vectors  fills up the five-dimensional graviton 
multiplet and the rest form a vector multiplet which contains a vector and 
five scalars. Thus, in this case we get a $\cN=4$ supergravity first discussed 
in \cite{C} coupled to a vector 
multiplet where the scalars parametrize $SO(1,1)\times 
SO(5,1)/SO(5)$ \cite{AT}. 

  The vectors of the five-dimensional graviton multiplet transform
in the $5+1$ rep. of the  $USp(4)$ automorphism 
group and it is not difficult to see by comparing
the Cern-Simons term of eq.(\ref{sts}) with the corresponding
term of the  $\cN=4$ $d=5$ supergravity \cite{AT} that the $USp(4)$ 
singlet is $B_{m}$. Thus, in  the bosonic part of the $\cN=4$ $d=5$
supergravity theory for the graviton multiplet we must 
also include  the four-derivative
interaction terms
\be
 S_{R^2}^{\cN=4}\sim \int d^5x\sqrt{-g}
\left(R_{mnrs}R^{mnrs}-\frac{1}{2}
\e^{mnpqr}B_{m}R_{abnp}
R_{abqr}\right)
%\!\!\!\!\!\!\!
%\phantom{\frac{{E^A}^A}{{E^A}^A}}\!\!\!\!\!\!\!\right)
\, , \nonumber 
\ee 
 Thus, unlike the maximal $\cN=8$ theory, the $\cN=4$ $d=5$ theory 
has four-derivative $R^2$ terms. These are one-loop terms and thus, subleading
with respect to the dominant two-derivative ones. The four-derivative
interactions exist in the ungauged theory and  
we expect to exist in the gauge one
as well. This is supported also from the CFT side as we will see later.

\subsection{The $\cN=2$ $d=5$ supergravity}

This theory has also been constructed in \cite{C} and it is very similar 
to eleven-dimensional supergravity. The five-dimensional action can be 
found by compactification of M-theory on a Calabi-Yau manifold.
Then, the reduced five-dimensional theory is described by
the $\cN=2$ supergravity coupled to $h_{(1,1)}-1$ vectors and $h_{(2,1)}+1$
hypermultiplets \cite{CF},\cite{AFN}. 
Since there are higher-derivative terms in M-theory given in eq.(\ref{M}),
one expects that there should be similar terms in the 
$\cN=2$ $d=5$ theory as well. For a 
Calabi-Yau compactification, we may express  the three-form 
of eleven-dimensional supergravity $C$ as $C=\sum_\Lambda
A_1^\Lambda\wedge \omega_\Lambda$ where $\Lambda=1,...,h_{(1,1)}$ and
$\omega_{\Lambda}$ are the corresponding $(1,1)$ harmonic forms 
on the $CY_3$. In this case,  an interaction of the form 
\be
S_5^L\sim \int \alpha_\Lambda A_1^\Lambda \wedge Tr R^2\, , 
\ee
where 
\be
\alpha_\Lambda=\int_{CY_3} 
\omega_\Lambda \wedge Tr R^2\, , \label{a}
\ee
is generated in five dimensions.
This term is actually the bulk term needed to cancel the anomalies due to 
wrapped fivebranes around the CY four-cycles \cite{FM}.
On the other hand, four-derivative terms can also be emerge, as has been
shown in $\cite{AF}$  by integrating the $\e_{11}\e_{11}R^4$ term in
eq.(\ref{M}). Indeed, the integration over the $CY_3$ produces the 
effective term 
\be
S_5'\sim \int {\e^{mnpq}}_{r}R_{mn}\wedge
R_{pq}\wedge dx^r=\frac{1}{2}\int d^5x \sqrt{-g} \left(R_{mnpq}
 R^{mnpq}+\ldots \right) \, . 
\ee
As a result, the ungauged $\cN=2$ five-dimensional supergravity
contains the four-derivative interactions
\be
{S_{R^2}^{\cN=2}}\sim  \int d^5x \sqrt{-g} \left( \beta R_{mnpq}
 R^{mnpq}+\alpha_\Lambda A_1^\Lambda\wedge Tr R^2\right)\, , \label{2}
\ee
where $\beta$ is proportional to $c_2\cdot k$ \cite{AF} 
with $c_2,\, k$ the second
Chern class and K\"ahler class of the $CY_3$,  respectively. In the gauged 
theory we expect that eq.(\ref{2}) survives consistently with the 
field theory expectations as will see below. The gauged U(1) theory  
have a one-form potential $B_1$ which  is a combination of the 
$h_{1,1}$ vectors of the theory \cite{GT} and the relative coefficients
of the two terms in eq.(\ref{2}) are related by supersymmetry in the same 
way that the R-current anomaly is related to the trace anomaly in the field
theory side.

\section{Field Theory Results }

We recall the expression for the trace anomaly of a four-dimensional 
conformal field theory in external gravitational field 
(see \cite{noi} for the notation),
\begin{equation}
\Theta={1\over 16\pi^2}\left[
c (W_{\mu\nu\rho\sigma})^2-a (\tilde R_{\mu\nu\rho\sigma})^2\right]+
{c\over 6\pi^2}(F_{\mu\nu})^2,
\label{ano}
\end{equation}
where $W_{\mu\nu\rho\sigma}$ and $R_{\mu\nu\rho\sigma}$ 
are the Weyl and Riemann tensors, respectively, while 
$F_{\mu\nu}$ is the field strength
of the $U(1)$ field $B_\mu$ coupled to the $R$-current. We recall that
$\mu, \nu\ldots=0,...,3$. In the free-field limit we have 
$c={1\over 24}(3N_v+N_\chi)$ and $a={1\over 48}(9N_v+N_\chi)$, where
$N_v$ and $N_\chi$ are the numbers of vector multiplets and
chiral multiplets, respectively.
$c$ and $a$ are marginally uncorrected \cite{noi}, so their values
are independent of the coupling constant in the theories that 
we are considering.
We can rewrite (\ref{ano}) as
\begin{equation}
\Theta={1\over 16\pi^2}\left[2(2a-c)R_{\mu\nu}R^{\mu\nu}+{1\over 3}
(c-3a)R^2 +(c-a)
R_{\mu\nu\rho\sigma}R^{\mu\nu\rho\sigma}\right]+
{c\over 6\pi^2}(F_{\mu\nu})^2.
\label{511}
\end{equation}
The factor $c-a$ multiplies the term containing the Riemann tensor and
this is one way to detect the subleading corrections. 
 
Supersymmetry relates the trace anomaly
to the $R$-current anomaly. Now, in $\cN$=1 supersymmetric theories,
there is only one such current, in general. It 
reads \cite{noi}
\[
R^{(\cN=1)}_\mu={1\over 2}\bar\lambda \gamma_\mu \gamma_5 \lambda
-{1\over 6}(\bar \psi \gamma_\mu\gamma_5\psi+
\widetilde{\bar \psi} \gamma_\mu\gamma_5\widetilde{\psi})+{\rm scalars},
\]
where $\lambda$ is the gaugino and $\psi,\tilde \psi$
are the matter fermions.
The anomaly formula is \cite{noi}
\begin{equation}
\partial_\mu(\sqrt{g}R^\mu)_{(\cN=1)}={1\over 24 \pi^2}(c-a)\,
\varepsilon_{\mu\nu\rho\sigma}
{R^{\mu\nu}}_{\beta\gamma}R^{\rho\sigma\beta\gamma}+
{1\over 9\pi^2}(5a-3c)F_{\mu\nu}\tilde F^{\mu\nu}.
\label{rcu0}
\end{equation}
On the other hand, in $\cN$=2 theories one has an $SU(2)\otimes U(1)$-group
of $R$-currents and the $U(1)$ $R$-current reads, in the notation
of \cite{n=2},
\[
R^{(\cN=2)}_\mu={1\over 2}
\bar\lambda_i \gamma_\mu \gamma_5 \lambda_i-
{1\over 2}(\bar \psi \gamma_\mu\gamma_5\psi+
\widetilde{\bar \psi} \gamma_\mu\gamma_5\widetilde{\psi})+{\rm scalars},
\]
where $\lambda_i$, $i=1,2$, are the two gauginos.
It satisfies \cite{n=2}
\begin{equation}
\partial_\mu(\sqrt{g}R^\mu)_{(\cN=2)}=-{1\over 8 \pi^2}(c-a)\,
\varepsilon_{\mu\nu\rho\sigma}
{R^{\mu\nu}}_{\beta\gamma}R^{\rho\sigma\beta\gamma}+
{3\over \pi^2}(c-a)F_{\mu\nu}\tilde F^{\mu\nu}.
\label{rcu}
\end{equation}
The above relationships provide
alternative ways to detect the subleading corrections to $c-a$,
and exhibit some difference between the $\cN$=1 and $\cN$=2 cases.

The divergence of the $R$-current couples to the longitudinal component of the
$U(1)$-field $B_\mu$. Indeed, taking $B_\mu$
to be pure gauge, $B_\mu=\partial_\mu\Lambda$, we have
\[
\int {\rm d}^4x\, \sqrt{g}R^\mu B_\mu\rightarrow -\int_{M} 
{\rm d}^4x\, 
\Lambda \partial_\mu(\sqrt{g}R^\mu).
\]
The external sources are viewed as boundary limits
of fields in five-dimensional supergravity.

The string-one-loop subleading correction derived from ten dimensions
reads in five dimensions
\be
\int_{\cal M} {\rm d}^5x\,\e^{mnpqr}B_m{R^a}_{bnp}{R^b}_{aqr}. \label{AA}
\ee
After replacing $A_m$ with $\partial_m\Lambda$, 
the boundary limit is
straightforward, in the sense that we do not need to use explicit 
Green functions. This is true in general for
anomalies, since it is
sufficient to look at the local part of the triangle diagram to 
reconstruct the full correlator.
In particular, in a conformal
field theory the three-point function $\langle R TT\rangle $ is unique
up to a factor and 
therefore uniquely determined by the anomaly we are considering.
We have
\[
-\int_{\cal M} {\rm d}^5x\,\e^{mnpqr}\partial_m
\left(\Lambda {R^a}_{bnp}{R^b}_{aqr}\right)
\rightarrow
-\int_{M} {\rm d}^4x\,\e^{\mu\nu\rho\sigma}
\Lambda{R^\alpha}_{\beta\mu\nu}{R^\beta}_{\alpha\rho\sigma}
\]
where $M=\partial{\cal M}$. The anomaly
correlator $\langle\partial R(x)
\,T_{\mu\nu}(y)\, T_{\rho\sigma}(z)\rangle $
is derived by taking one functional derivative 
with respect to $\Lambda$ and two functional derivatives with
respect to the metric tensor. The result can be written in the form of an
operator equation
\[
\partial_\mu(\sqrt{g}R^\mu)=f\,\varepsilon_{\mu\nu\rho\sigma}
{R^{\mu\nu}}_{\beta\gamma}R^{\rho\sigma\beta\gamma}.
\]
for some factor $f$. Quantum field theory, formula (\ref{rcu}),
says that $f={1\over 24 \pi^2}(c-a)$ for $\cN$=1
and $f=-{1\over 8 \pi^2}(c-a)$ for $\cN$=2. String theory gives the geometrical
interpretation of this number via formula (\ref{a}).

Similar remarks can be repeated for the contribution 
$R_{\mu\nu\rho\sigma}R^{\mu\nu\rho\sigma}$ in the trace anomaly (\ref{511}).
Instead, the coefficient of the term
$F_{\mu\nu}\tilde F^{\mu\nu}$ in the $R$-current anomalies 
presents two different behaviours: it is leading for 
$\cN$=1 and subleading for $\cN$=2.
The string/supergravity description is in agreement with 
this fact (see below), which is a nontrivial cross-check
of the consistency of
our picture and, as a bonus, provides
a precise prediction for some string-loop corrections.

\subsection{$\cN=4$}

The coefficient  $f$ is subleading in the large $N$ limit
and identically zero in $\cN=4$ supersymmetric Yang-Mills theory.
With $G=SU(N)$, we have $c=a={1\over 4}(N^2-1)$ so that $f=0$.
Since the supergravity dual of this theory is 
the $\cN=8$ $d=5$ gauge supergravity,  
we do not have four-derivative interactions. Indeed, as we have discussed 
in section 2, there are no such interactions as a result of 
the toroidal compactification, consistently with the field-theory 
expectations.  

\subsection{$\cN=2$}

In general in this case $c$ and $a$ is not exactly equal. Consider 
for example the $\cN=2$ finite theory with $G=SU(N)\otimes SU(N)$
and two copies of hypermultiplets in the $R=(N,\bar N)$ representation.
We have
\[
c={1\over 2}N^2-{1\over 3},~~~~~~~~~
a={1\over 2}N^2-{5\over 12},
~~~~~~~~~~~~~~~
c-a={1\over 12}.
\]
The supergravity dual of this theory has been found to be type IIB 
theory on $AdS_5\times S^5/Z_2$ \cite{KS},\cite{KW} and thus it is 
$\cN=4$ $d=5$ gauged supergravity. The latter has, as we have seen 
in section 3, $R^2$ interactions of precisely the correct form to 
account for R-current anomaly eq.(\ref{rcu}) and the 
$R_{\mu\nu\rho\sigma}R^{\mu\nu\rho\sigma}$-term
in (\ref{511}).

The $U(1)_R^3$ term
$F_{\mu\nu}\tilde F^{\mu\nu}$ is subleading, as we see in (\ref{rcu}).
Thus we expect that a term of the form $B\wedge F\wedge F$ should exist
in the low-energy string effective action. The absence of this term
in the five-dimensional $\cN=4$ leading supergravity action has been noticed 
in \cite{FZ}\footnote{We thank S. 
Ferrara for clarifying discussions on this point}. However,
as we see here, this term is actually subleading and should come from
the string one-loop computation.

\subsection{$\cN=1$}

As in the previous case, here  we have also, in general, non-vanishing $c-a$.
One can  take for example the $\cN=1$ 
theory with $G=SU(N)\otimes SU(N)\otimes SU(N)$ 
and three copies of
$R=(N,\bar N,1)\oplus {\rm cyclic\, perm.s}$ \cite{KS}. We have
\[
c={3\over 4}N^2-{3\over 8},~~~~~~~~~a={3\over 4}N^2-{9\over 16},
 ~~~~~~~~~~~c-a={3\over 16}.
\]
The supergravity dual is the  $\cN=2$ $d=5$ gauged supergravity theory.
As above, this theory has the correct $R^2$ terms eq.(\ref{2}) 
to produce  the R-current anomaly.   

The $U(1)_R^3$ term
$F_{\mu\nu}\tilde F^{\mu\nu}$ is leading, by formula (\ref{rcu0}),
and the corresponding bulk term $B\wedge F\wedge F$ is indeed
present in the supergravity Lagrangian \cite{GT},\cite{FZ}.

\subsection{Interpretation at the level of quantum conformal algebras.}

The quantum field theoretical origin of the subleading corrections
to $c-a$ was explained in \cite{n=2}. One can study the OPE
of conserved currents, or in general finite operators, which generate
a hierarchy of higher spin tensor currents organized into
supersymmetric multiplets. In the $\cN$=2 case 
there is one pair of current multiplets for each even spin 
and one current multiplet for each odd-spin.
All multiplets have length 2 in spin units.
In $\cN=4$ \cite{anom}, instead, there is one 4-spin-long multiplet
for each even spin, plus the stress-tensor.
Some powerful theorems
in quantum field theory imply that a relevant part
of the hierarchical 
structure is preserved to all orders in the coupling constant
(see \cite{anom,n=2} for details)
and therefore should be visible 
around the strongly coupled limit.

In particular, the
$\cN=2$ algebra contains 
a current multiplet ${\cal T}^*$ that mixes with the 
multiplet ${\cal T}$ of the stress-tensor. 
The mixing is 
responsible of the desired effect.
We recall here the argument.

The central charges $c$ and $a$ are encoded, as we see from (\ref{ano}),
into the three-point function
$\langle T(x)T(y)T(z)\rangle $, which we can study 
by taking the $x\rightarrow y$ limit and using the operator product expansion
of \cite{n=2}, written schematically as
$T(x)T(y)=\sum_n \hbox{$c_n(x-y)$}{\cal O}_n\left({x+y\over 2}\right)$.
The operators
${\cal O}_n$ that mix with the stress-tensor
contribute via the two-point functions
$\langle {\cal O}_n\,T\rangle$.
In the $\cN=2$ quantum conformal algebra 
these are the stress-tensor $T_{\mu\nu}$
itself and a second operator $T^*_{\mu\nu}$, which 
in the free-field limit
is proportional to 
$T_v-2T_m$, $T_v$ and $T_m$ being the vector-multiplet
and hypermultiplet contributions to the stress-tensor.
${\cal O}_n=T$ produces a contribution
$\langle TT\rangle$, which is leading and actually equal to $c$.
The contribution
from ${\cal O}_n={\cal T}^*$,
$\langle {\cal T}^*\, {\cal T}\rangle$,
is non-vanishing and subleading,
precisely ${\cal O}(1)$.
Apart from this remnant,
${\cal T}^*$, as well as the other non-conserved 
operators ${\cal O}_n$, decouple from the theory
in the strongly coupled large-$N$ limit.

The identification between string corrections and ${\cal T}^*$
is a step towards 
the reconstruction of the spin hierarchies of \cite{anom,n=2}
as string excitations around the supergravity limit.
We expect that the full hierarchies of \cite{anom,n=2} can be 
found in the string description. 
Roughly, the vocabulary should be as follows.
i) Higher-spin current multiplets 
that are orthogonal to the stress-tensor correspond
to $\alpha^\prime$-corrections. ii) The renormalization 
mixing between pairs of current multiplets is mapped onto
$1/N$-corrections. 

There is however an effect that is not included in this classification.
$c$ and $a$ receive, separately, a $1/N$-correction that
is not detectable via the difference $c-a$ and is present also
in $\cN=4$
(where no renormalization mixing takes place \cite{anom}).
The identification of the string origin of this
kind of subleading corrections is still missing.

We recall, finally, that there are theories in which $c-a$ is leading.
They have a quite different quantum conformal algebra \cite{n=2} and
the difference persists in the closed (large $N$,
large $g^2N$) limit. At the moment,
there is no string interpretation to this more general class
of conformal field theories in four dimensions.

\vspace{.5cm}

{\it Acknowledgement:} We would like to thank S. Ferrara and E. Kiritsis 
for extensive discussions.


\begin{thebibliography}{99}

\bibitem{malda} J. Maldacena,  Adv. Theor. Math. Phys. 2 (1998)231,
hep-th/9711200.
\bibitem{kleb} S.S. Gubser, I.R. Klebanov and A.M. Polyakov, Phys. Lett. 
B 428 (1998)105,   hep-th/9802109.
\bibitem{wit1} E. Witten, Adv. Theor. Math. Phys.2 (1998)253, 
hep-th/9802150; Adv. Theor. Math. Phys. 2 (1998)505, hep-th/9803131. 
%%%%%
%\bibitem{LNV} A. Lawrence, N. Nekrasov and  C. Vafa,  
 %  hep-th/9803015.
%\bibitem{BJ} M. Bershadsky, Z. Kakushadze and  C. Vafa, 
%hep-th/9803076; M. Bershadsky, A. Johansen, hep-th/9803249.
%\bibitem{Kze} Z. Kakushadze, hep-th/9803214, hep-th/9804184.
 
%\bibitem{IMSY} N. Itzhaki, J.M. Maldacena,
%J. Sonnenschein and  S. Yankielowicz, hep-th/9802042.
\bibitem{FF} S. Ferrara and C. Fronsdal, Phys. Lett. B 433 (1998)19,
hep-th/9802126;  Class. Quant. Grav. 15 (1998)2153, hep-th/9712239;\\
 C. Fronsdal, S. Ferrara and A. Zaffaroni, Nucl. Phys. B 532(1998)153,  
hep-th/9802203.
\bibitem{FZ} S. Ferrara and A. Zaffaroni, Phys. Lett. B431 (1998)49, hep-th/9803060.
\bibitem{GRW} M. G\"unaydin, L.J. Romans and N.P. Warner, 
Nucl. Phys. B 272 (1986)598; Phys. Lett. B 154 (1985)268.
\bibitem{Ro} L.J. Romans, Nucl. Phys. B 267 (1986)433.
\bibitem{GT} M. G\"unaydin, G. Sierra and P.K. Townsend, Nucl. Phys. B
253 (1985)573.
\bibitem{noi}  D. Anselmi, D.Z. Freedman, M.T. Grisaru and A.A. Johansen,
Nucl. Phys. B 526 (1998)543, hep-th/9708042. 
\bibitem{HS} M. Henningson and K. Skenderis,
J. High Energy Phys. 9807 023 (1998), hep-th/9806087. 
\bibitem{Gu} S.S. Gubser, hep-th/9807164. 
\bibitem{n=2} D. Anselmi,  hep-th/9811149.
\bibitem{anom} D. Anselmi, hep-th/9809192.
\bibitem{GZ} M.T. Grisaru, A.E.M. van de Ven and D. Zanon, Nucl. Phys.
B 277 (1986)388.
\bibitem{GW} D.J. Gross and E. Witten, Nucl. Phys. B 277 (1986)1.
\bibitem{FT} E.S. Fradkin and  A.A. Tseytlin,  Nucl. Phys. B 227 (1983)252.
\bibitem{GV} M.B. Green and  P. Vanhove, Phys. Lett. B 408 (1997)122,
 hep-th/9704145. 
 \bibitem{S} J. Schwarz, Phys. Rep. 89 (1982) 223.
\bibitem{TT} A.A. Tseytlin, Nucl. Phys. B 467 (1996)383, hep-th/9512081.
\bibitem{KP} E. Kiritsis and B. Pioline, Nucl. Phys. B 508 (1997)509,
 hep-th/9707018. 
\bibitem{BR} E. Bergshoeff and M. de Roo, Nucl. Phys. B  328 (1989)439;\\
M. de Roo, H. Suelmann and A. Wiedmann, Phys. Lett. B 280 (1992)39; Nucl. 
Phys. B 405 (1993)326.
\bibitem{GG} M.B.  Green and M. Gutperle, Nucl. Phys. B 498 (1997)195, 
hep-th/9701093.
\bibitem{KP1} A. Kehagias and H. Partouche, Phys. Lett. B 422 (1998)109, 
hep-th/9710023.
\bibitem{RT} J.G. Russo and  A.A. Tseytlin, Nucl. Phys. B 508 (1997)245,
hep-th/9707134.
\bibitem{KP2} A. Kehagias and H. Partouche, 
Int. J. Mod. Phys. A13 (1998)5075, hep-th/9712164.
\bibitem{DM} M.J. Duff,  J.T. Liu and R. Minasian, Nucl. Phys. B 452
(1995) 261. 
 \bibitem{C} E. Cremmer, in ``{\it Superspace and Supergravity''}, ed. S.W. 
Hawking and M. Rocek (CUP, 1981).
\bibitem{GZP} L. Girardello, M. Petrini, M. Porrati and A. Zaffaroni,
 hep-th/9810126.
\bibitem{DZZ}  J. Distler and  F. Zamora, hep-th/9810206. 
\bibitem{PWW} A. Khavaev, K. Pilch and  N.P. Warner,  hep-th/9812035. 
\bibitem{BG} T. Banks and M.B. Green, J. High Energy Phys. 5 (1998)2, 
hep-th/9804170.
\bibitem{KKP} A. Gregori, E. Kiritsis, C. Kounnas, N.A. Obers, 
P.M. Petropoulos and B. Pioline,  Nucl. Phys. B 510 (1998)423, 
hep-th/9708062. 
\bibitem{R} L.J. Romans, Nucl. Phys. B 269 (1986)691.
\bibitem{AT} M. Awada and P. Townsend, Nucl. Phys. B 255 (1985)617. 
 \bibitem{CF} A.C. Cadavid, A. Ceresole, R. D'Auria
and  S. Ferrara, Phys. Lett. B 357 (1995)76,  hep-th/9506144.
 \bibitem{AFN}  I. Antoniadis, S. Ferrara  and  T.R. Taylor,
 Nucl. Phys. B 460 (1996)489. 
\bibitem{FM} S. Ferrara, R.R. Khuri and R. Minasian, Phys. Lett. B 375 
(1996)81.
\bibitem{AF} I. Antoniadis, S. Ferrara, R. Minasian and  K.S. Narain,
 Nucl. Phys. B 507 (1997)571, hep-th/9707013.
 \bibitem{KS} S. Kachru and E. Silverstein,  
Phys. Rev. Lett. 80 (1998)4855,  hep-th/9802183.
\bibitem{KW} I. Klebanov and E. Witten, hep-th/9807080. 








%\bibitem{nach} O. Nachtmann, Positivity constraints for
%anomalous dimensions, Nucl. Phys. B 63 (1973) 237.

%\bibitem{ooguri} G.T. Horowitz and H. Ooguri,  hep-th/9802116.

%\bibitem{FFZ} C. Fronsdal, S. Ferrara and A. Zaffaroni, hep-th/9802203;
%S. Ferrara and A. Zaffaroni,  hep-th/9803060; S. Ferrara, M.A. 
%Lledo and A. Zaffaroni, hep-th/9805082.
%\bibitem{FKZ} S. Ferrara, A. Kehagias, H. Partouche and A. Zaffaroni, 
% hep-th/9803109; hep-th/9804006.
%\bibitem{HS} M. Henningson, K. Sfetsos, hep-th/9803251.  
%\bibitem{sjr} S.J. Rey and  J. Yee,  
%hep-th/9803001; S.J. Rey, S. Theisen and  J.T. Yee,  hep-th/9803135;
%J.M. Maldacena,  hep-th/9803002.
%J.A. Minahan,  hep-th/9803111;J.A. Minahan and N.P. Warner, hep-th/9805104.
%\bibitem{BRAND} A. Brandhuber, N. Itzhaki, J. Sonnenschein, S. Yankielowicz,
% hep-th/9803137.
%\bibitem{AOY} O. Aharony, Y. Oz and Z. Yin, hep-th/9803051.


%\bibitem{arefa} I.Ya. Aref'eva and I.V. Volovich,  hep-th/9803028; 
%hep-th/9804182.

%\bibitem{behr} E. Bergshoeff and  K. Behrndt,  hep-th/9803090.
%\bibitem{lu} M.J. Duff, H. Lu and C.N. Pope,  hep-th/9803061.
 


%\bibitem{berkooz} M. Berkooz,  hep-th/9802195; J. Gomis,  hep-th/9803119;
%%Y. Oz and  J. Terning,  hep-th/9803167;
%E. Halyo,  hep-th/9803193;
%F. Brandt, J. Gomis and J. Simon,  hep-th/9803196; R. Entin and J. Gomis, 
%hep-th/9804060; R. Emparan, hep-th/9804031.


%\bibitem{aha} 
%R.G. Leigh and M. Rozali,  hep-th9803068;
%S. Minwalla,  hep-th/9803053;
%E. Halyo,  hep-th/9803077.


%\bibitem{horo} G.T. Horowitz and S.F. Ross,  hep-th/9803085.
%\bibitem{BB} E. Bergshoeff and K. Behrndt,  hep-th/9803090.
%\bibitem{ITY} N. Itzhaki, A.A. Tseytlin and S. Yankielowicz,  hep-th/9803103.

%\bibitem{AND} L. Castellani, A. Ceresole, R. D'Auria, S. Ferrara, 
%P. Fr\'e and M. Trigiante, hep-th/9803039;   L. Andrianopoli and S. Ferrara,  %hep-th/9803171.

%\bibitem{VOLO} I.V. Volovich,  hep-th/9803174,  hep-th/9803220. 
%\bibitem{AOT} C. Ahn, K. Oh and R. Tatar, hep-th/9804093.
%\bibitem{FMMR} D.Z. Freedman, S.D. Mathur, A. Matusis and L. Rasteli, 
%hep-th/9804058.
%\bibitem{FS} A. Fayyazuddin and M. Spali\'nski, hep-th/9805096.
%\bibitem{wit2} E. Witten, hep-th/9805112.
%\bibitem{SW} L. Susskind and E. Witten, hep-th/9805114.
%\bibitem{WN} W. Nahm, Nucl. Phys. B 135 (1978)149.

%\bibitem{Sw} J.H. Schwarz and P. West, Phys. Lett. B 126 (1983)301; 
%P.S. Howe and P. West, Nucl. Phys. B 238 (1984)181; J.H. Schwarz, 
%Nucl. Phys. B 226 (1983)269.


%\bibitem{DP} M.J. Duff and J.X. Lu, Phys. Lett. B 273  (1991)409.
%\bibitem{HSt} G.T. Horowitz and A. Strominger, Nucl. Phys. B 360 (1991)197.
%\bibitem{PW} C.N. Pope and N.P. Warner, Class. Quantum Grav. 2 (1985)L1.
%\bibitem{ALB} A.L. Besse, {\it ``Einstein Manifolds''}, Springer-Verlag, 1987.

%\bibitem{AF} A. Futaki, {\it ``K\"ahler-Einstein Metrics and Integral
%Invariants''}, Springer-Verlag, 1988. 
%\bibitem{YTS} Y.T. Siu, Ann. Math. 127 (1988)585.  
%\bibitem{TY} G. Tian and S.T. Yau, Commun. Math. Phys. 112 (1987)175.

%\bibitem{T1} G. Tian, Invent. Math. 89 (1987)225; 101 (1990)101; 130 (1997)1. 
%\bibitem{Vafa} C. Vafa, 
%Nucl. Phys. B 469 (1996)403.

%\bibitem{DW} R. Donagi, A. Grassi and E. Witten, 
%Mod. Phys. Lett. A11 (1996), hep-th/9607091.
%\bibitem{CL} G. Curio and D. L\"ust, 
%Int. J. Mod. Phys. A12 (1997)5847, hep-th/9703007.





\end{thebibliography}
\end{document}